\documentstyle[aps,epsfig]{revtex}  % Specifies the document style.
\newcommand{\be}{\begin{equation}}
\newcommand{\ee}{\end{equation}}

\begin{document}
\title{Bose Glass in Large $N$ Commensurate Dirty Boson Model}
\author{M. B. Hastings}
\address{Physics Department, Jadwin Hall\\
Princeton, NJ 08544\\
hastings@feynman.princeton.edu}
\maketitle
\begin{abstract}
The large $N$ commensurate dirty boson model, in both
the weakly and strongly commensurate cases, is considered
via a perturbative renormalization group treatment.
In the weakly commensurate case, there exists a fixed line
under RG flow, with varying amounts of disorder along the line.
Including $1/N$ corrections causes the system to flow to strong disorder,
indicating that the model does not have
a phase transition perturbatively connected to the 
Mott Insulator-Superfluid (MI-SF)
transition.  I discuss the qualitative effects of
instantons on the low energy density of excitations.
In the strongly commensurate case, a fixed point found previously
is considered and results are obtained for 
higher moments of the correlation functions.
To lowest order, correlation functions have a log-normal distribution.
Finally, I prove two interesting theorems for large $N$ vector models 
with disorder, relevant to the problem of replica symmetry breaking and
frustration in such systems.
   \end{abstract}
\section{Introduction}
The dirty boson problem, a problem of repulsively interacting bosons in
a random potential, has been the subject of much theoretical
work\cite{dirtyboson}.  In the
zero temperature quantum problem, the system can undergo a phase
transition between a Bose glass phase and a superfluid phase.
An action that may be used to describe this transition is
\be
\label{orig}
\int d^dx \, dt \, \Bigl(
+\partial_{x} \overline \phi(x,t) \partial_{x} \phi(x,t)
+\partial_{t} \overline \phi(x,t) \partial_{t} \phi(x,t)
+w(x)\overline\phi(x,t)\partial_t\phi(x,t) -
U(x) \overline\phi\phi +g(\overline\phi\phi)^2 \Bigr)
\ee

This action describes a system with a number of phases.  In the pure case,
in which $U(x)$ is a constant, and $w(x)$ vanishes everywhere (the commensurate
case), there is a phase transition from a gapped Mott insulator phase
to a superfluid phase as $U$ increases.  If we consider a 
case in which $U(x)$ is not constant,
but $w(x)$ still vanishes everywhere, there will be a sequence of transitions
from a gapped Mott insulator, to a gapless (but with exponentially vanishing
low energy density of states) Griffiths phase, and then to
a superfluid phase.  The Griffiths phase occurs due to the possibility of
large, rare regions in which fluctuations in $U(x)$ make the system locally
appear superfluid.  

If we consider the case in which both $U(x)$ and $w(x)$ are fluctuating, 
there will also appear a Bose glass phase in which there
is a density of states tending to a constant at low energy and an infinite 
superfluid susceptibility\cite{boseglass}.  The physical basis for this
phase is the existence of localized states, in which a competition between
chemical potential and repulsion causes the system to desire a certain
number of particles to occupy each localized state.  There exist excitations
involving adding or removing one particle from these states, and these
excitations lead to the diverging susceptibility.  However, it is clear that
the Bose glass phase is very similar to the Griffiths phase, in that both
involve regions of finite size.  In the Griffiths phase one needs regions
of arbitrarily large size, while in the Bose glass phase one only needs
regions large enough to support a localized state, with a nonvanishing
number of particles occupying that state, in order to produce the diverging
susceptibility characteristic of the phase.  In a system in which the
disorder is irrelevant at the pure fixed point, so that the fluctuations in
$U(x)$ and $w(x)$ scale to zero, one will still find a Bose glass phase
as there will, with low probability, exist regions that can give rise to
these localized states.  Thus, the interesting question to answer is not
whether the Bose glass phase exists, but whether there exists a fixed point
at which fluctuations in $w(x)$ are weak, so that the critical exponents are
near those of the pure MI-SF transition.  The most likely alternative 
would be governed by the scaling theory of Fisher et. al.,
which has very different critical exponents\cite{dirtyboson,herbut}.  We will
refer to this scaling theory as the phase-only transition, as one
assumes fluctuations in the amplitude of the order parameter are
irrelevant at the critical point.

Recently, a large $N$ generalization of equation (\ref{orig}) was considered
in the restricted case $w(x)=0$\cite{me}.  We will refer to this
case as the strongly commensurate case, while the situation in
which $w(x)$ vanishes on average, but has nonvanishing fluctuations
will be known as the weakly commensurate case.  We consider a system defined
by the partition function
\be
\label{orN}
\int \Bigl(\prod\limits_{x,t}\delta(\overline\phi_i(x,t)\phi^i(x,t)-
N\sigma^2(x))\Bigr) [d\phi_i] e^{-S}
\ee
where
\be
S=\int d^dx \, dt
(\partial_{x} \overline \phi_i(x,t) \partial_{x} \phi^i(x,t)+
\partial_{t} \overline \phi_i(x,t) \partial_{t} \phi^i(x,t)
+w(x)\overline\phi_i(x,t)\partial_t\phi^i(x,t))
\ee
Here, we have, for technical simplicity later, replaced the quartic
interaction by a $\delta$-function.  For most of the paper, a $\delta$-function
interaction will be used.  However, for generality in the last
section, we will return to quartic interactions.
The disorder in $U(x)$ will be replaced, in the $\delta$-function case,
by weak fluctuations in $\sigma^2$.  We will consider $\sigma^2=
\sigma_0^2+\delta\sigma^2$ where $\sigma_0^2$ is a constant piece used to
drive the system through the phase transition and $\delta\sigma^2$ is
a fluctuating piece.

The advantage of the large $N$ formulation of the problem is that
for any fixed realization of the problem one may exactly solve by
system by finding the solution of the self-consistency equation
\be
\label{sc}
\sigma^2(x)=\langle x,t=0|(-\partial_{x}^2-\partial_t^2+w(x)\partial_t 
+\lambda(x))^{-1}|x,t=0\rangle
\ee
or
\be
\sigma^2(x)=\int d\omega \,
\langle x,t=0|(-\partial_{x}^2+\omega^2+i w(x)\omega 
+\lambda(x))^{-1}|x,t=0\rangle
\ee
where $\lambda(x)$ is a Lagrange multiplier field for enforce the 
$\delta$-function constraint on the length of the spins.  
After solving the
self-consistency equation, any correlation function can be found simply
by finding the Green's function of a non-interacting field $\phi$ with
action $\int d^dx \, dt \, \overline\phi (-\partial_{x}^2-\partial_t^2
+w(x)\partial_t+\lambda(x))\phi$.

In equation (\ref{sc}), we assume that the Green's function on the right-hand
side has been renormalized by subtracting a divergent quantity.  Specifically, 
we will take a Pauli-Villars regularization for the Green's function, 
and take the regulator mass to be very large, while adding an appropriate 
divergent constant to $\sigma^2$ on the left-hand side.  The cutoff for the 
regulator is completely different from the cutoff for fluctuations in 
$\delta\sigma^2$ that will be used for the RG later; the cutoff for the
regulator will be much larger than the cutoff for fluctuations in 
$\delta\sigma^2$ and $w(x)$, and will be unchanged under the RG.

In the previous work, exact results were obtained for the
critical exponents for average quantities.  In the present paper, I
will first consider the problem in which $w(x)\neq0$, although $w(x)$ vanishes
on average (the weakly commensurate dirty boson problem).  I will
consider general problems of the large $N$ system with terms linear in
the time derivative.  Next, a lowest order perturbative RG treatment
will be used to consider critical behavior.  Instanton corrections to
the perturbative treatment will be briefly discussed after that.
Returning to the strongly commensurate case, previous results on the
fixed point will be extended to give results for higher moments of
the correlation functions.
Finally, as a technical aside, we consider the large $N$ self-consistency
equation in frustrated systems, and demonstrate that the
self-consistency equation always has a unique solution, as well as considering
the number of spin components needed to form a classical ground
state in frustrated systems.
\section{Bose Glass in the Large $N$ Limit}
Consider the following simple $0+1$ dimensional problem, at
zero temperature ($\beta\rightarrow\infty$):
\be
\int [d\phi_i(t)] \prod\limits_t\delta(\overline\phi_i(t)\phi^i(t)-N\sigma^2)
e^{\int_{-\beta/2}^{\beta/2}
\{ -\partial_t\overline\phi_i\partial_t\phi^i+A
\overline\phi_i\partial_t\phi^i \} dt}
\ee
The solution of this problem in the large $N$ limit via a self-consistency
equation requires finding a $\lambda$ such that
\be
\sigma^2=\int \frac{d\omega}{2\pi} \frac{1}{iA\omega+\omega^2+\lambda}
\ee
Then, by contour integration, for $\lambda>0$, we find 
$\sigma^2=\frac{1}{\sqrt{A^2+4\lambda}}$.  Then, 
\be
\lambda=\frac{1}{4}(\frac{1}{\sigma^4}-A^2)
\ee
Unfortunately, this result for $\lambda$ leads to $\lambda$ becoming negative 
for sufficiently large $A$.  Although perturbation theory will not see a 
problem, this is the signal for the Bose glass phase.  One must separate the 
self-consistency equation into two parts, one containing an integral over 
non-zero $\omega$, and one containing the term for zero $\omega$.  One finds 
(for finite $\beta$)
\be
\sigma^2=\int \frac{d\omega}{2\pi} \frac{1}{iA\omega+\omega^2+\lambda}+\frac{1}{\beta}\frac{1}
{\lambda}
\ee
Then, the self-consistency equation can always be solved using positive 
$\lambda$, but in the zero-temperature limit of the problem one will find 
that one needs $\lambda$ to be of order $\frac{1}{\beta}$, and at 
zero-temperature, there will appear a zero energy state.

Considering the original statistical mechanics problem of equations 
(\ref{orig},\ref{orN}), one will expect to see some non-zero
density of these zero energy states, indicating the presence of a gapless 
Bose glass phase, with diverging superfluid susceptibility.  Even if the 
fluctuations in $w(x)$ vanish at the fixed point, when the critical exponents 
are unchanged from the strongly commensurate problem, such zero energy states 
will exist as Griffiths effects leading to the appearance of
a Bose glass phase near the superfluid transition.

To perform a renormalization group treatment of the model, we will first 
proceed in a perturbative fashion in the next section, ignoring such zero 
energy effects.  For small fluctuations in $w(x)$, these zero energy states 
will be exponentially suppressed as will be
discussed in the section after that.
\section{Perturbative RG}
We will follow the RG techniques used in previous work on the large
$N$ problem\cite{me}.  We will work near $2+1$ dimensions, specifically
we will have $d=2+\epsilon$ space dimensions and 1 time dimension.  We
will at first work to one-loop in perturbation theory, which will, in
some cases, correspond to lowest order in $\epsilon$.  Some results
will be extended to all orders.  A fixed line is found for the large $N$
system.  The fixed points are destroyed by
$1/N$ corrections.

The RG is defined as follows: start with a system containing fluctuations
in $\delta\sigma^2$ and $w(x)$ up to some wavevector $\Lambda$.  Remove the
high wavevector fluctuations in these $\delta\sigma^2$ and $w(x)$ to
obtain a new system, with renormalized gradient terms 
$\partial_x^2$ and $\partial_t^2$ in the action, as well as renormalized
low wavevector $\delta\sigma^2$ and $w(x)$ terms.  Do this procedure so
as to preserve the average low momentum Green's function, as well
as the low wavevector fluctuations in $\lambda(x)$.  See previous work\cite{me}
for more details.

If we are working in $2+\epsilon$ space dimensions, and 1 time dimension,
we can easily work out the naive scaling dimensions of the disorder.  One
finds that if we assume Gaussian fluctuations in the disorder, with
\be \langle\delta\sigma^2(p)\delta\sigma^2(q)\rangle=(2\pi)^2\delta(p+q)S\ee
\be \langle w(p)w(q)\rangle=(2\pi)^2\delta(p+q)W\ee
then $S$ scales as length to the power $d-2=\epsilon$, while
$W$ scales as length to the power $2-d=-\epsilon$.  So, for $d>2$
we find that disorder in $\sigma^2$ is relevant at the pure fixed point,
while disorder in $w(x)$ is irrelevant.  For $d<2$ this is reversed.

Previously, a lowest order in $\epsilon$ calculation\cite{me} considering
only disorder in $\delta\sigma^2$ gave the following results.  For a given
problem, with fluctuations in $\delta\sigma^2$ up to wavevector $\Lambda$,
and self-consistency equation 
\be
\label{sc1}
\sigma^2(x)=\int d\omega \,\langle x,t=0|
\Bigl(-\partial_{x}^2+\omega^2+\lambda(x)\Bigr)^{-1})|x,t=0\rangle
\ee
one could define another problem, with fluctuations in $\delta\sigma^2$
only up to $\Lambda-\delta\Lambda$, with self-consistency equation
\be
\label{nsc}
(1-\frac{\delta\Lambda}{\Lambda}c_3L)
\sigma^2(x)=\int d^{(D-d)}\omega \,\langle x,t=0|
\Bigl(-(1+\frac{\delta\Lambda}{\Lambda}c_2L)
\partial_{x}^2+
(1+\frac{\delta\Lambda}{\Lambda}c_3L)
(\omega^2
+\lambda(x))\Bigr)^{-1})|x,t=0\rangle
\ee
such that the Green's function computed from the second self-consistency
equation agrees with the Green's function computed from the first 
self-consistency equation averaged over disorder at large wavevector.
Here we define $L=c_1^2\Lambda^{8-2D}S$ where
$c_1=\frac{1}{\pi}^{D/2} \frac{\Gamma(D-2)} {\Gamma(2-D/2)\Gamma^2(D/2-1)}$,
$c_2=(1-4/d)c_3$, and $c_3=2\frac{\pi^{d/2}}{\Gamma(d/2)}\Lambda^{d-4}$.

The results above, to one loop, were obtained by considering the large
wavevector fluctuations in $\lambda$ due to the large wavevector fluctuations
in $\delta\sigma^2$, and then finding how they renormalize the self-energy
and vertex.  To lowest order, one obtains the fluctuations in $\lambda$
by inverting a polarization bubble.  That is, one expands the self-consistency
equation to linear order in $\lambda$ to solve for large wavevector
fluctuations in $\lambda$ as a function of fluctuations in $\delta\sigma^2$.
One finds then that 
\be
\label{pol}
\delta\sigma^2(p)=c_1^{-1}p^{D-4}\lambda(p)+...
\ee
From this, we obtain fluctuations in $\lambda$ at wavevector $\Lambda$
which, to lowest order, are Gaussian with mean-square $L$.  See
figures 1,2, and 3.
For more details, see previous work\cite{me}.

It may easily be seen that, to lowest order, the addition
of the term $w(x)$ does not produce any additional large wavevector
fluctuations in $\lambda$, as equation (\ref{pol}) is still true to
lowest order in $w(x)$ and $\lambda(x)$.  However, the term $w(x)$ can produce 
a renormalization of the self-energy.  See figure 4.  The result is
to produce a term in the self-energy equal to
\be
\Sigma(p,\omega)=
-\delta\Lambda\omega^2\int\limits_{k^2=\Lambda^2}d^{d-1}k
\frac{1}{(p+k)^2+\omega^2}W
\ee
This is equal to
\be
-\frac{\delta\Lambda}{\Lambda}\omega^2Wc_4+...
\ee
where $c_4=\Lambda^{d-2}2\frac{\pi^{d/2}}{\Gamma(d/2)}$.

There is one other term that must be included in the RG flow equations
at this order.  The fluctuations in $\lambda$ due to the fluctuations
in $\delta\sigma^2$ can renormalize the vertex involving $w(x)$.  See
figure 5.
This will change the term $w(x)\partial_t$ in the self-consistency
equation to $w(x)\partial_t (1+\frac{\delta\Lambda}{\Lambda}c_3L)$.
Note that the renormalization of the $w(x)$ term is equal to the renormalization
of the $\omega^2$ and $\lambda(x)$ terms in the self-consistency equation.

Putting all the terms together, we find that with a lowered cutoff
$\Lambda-\delta\Lambda$, the renormalized theory is described by the
new self-consistency equation
\be
(1-\delta_3) \sigma^2(x)=
\int d\omega \,\langle x,t=0|
\Bigl(-(1+\delta_2)\partial_{x}^2+(1+\delta_4)\omega^2
+i(1+\delta_3) w(x)\omega
+(1+\delta_3) \lambda(x)\Bigr)^{-1})|x,t=0\rangle
\ee
where $\delta_3=\frac{\delta\Lambda}{\Lambda}c_3L$,
$\delta_2=\frac{\delta\Lambda}{\Lambda}c_2L$, and 
$\delta_4=\delta_3+\frac{\delta\Lambda}{\Lambda}Wc_4$.
Rescaling $\omega$ by $(1+\frac{\delta_2-\delta_4}{2})$ to make the 
coefficients in front of the
$\omega^2$ and $\partial_x^2$ terms the same, rescaling $\lambda$,
and then rescaling the spatial
scale to return the cutoff to $\Lambda$ we find
\be
\tilde\sigma^2(x)=\int d\omega \,\langle x,t=0|
\Bigl(-\partial_{x}^2+\omega^2+
i\tilde w(x)\omega+
\lambda(x)\Bigr)^{-1})|x,t=0\rangle
\ee
where
\be
\tilde \sigma^2=
(1-\delta_3+\delta_2+\frac{\delta_4-\delta_2}{2}+(1+\epsilon)
\frac{\delta\Lambda}{\Lambda}) \sigma^2(x)
\ee
\be
\tilde w(x)=
i(1+\delta_3-\delta_2+\frac{\delta_2-\delta_4}{2}) w(x)
\ee
From this, we extract RG flow equations for $\sigma_0^2$, $S$, and $W$.
The result is
\be
\frac{d{\rm ln}\sigma_0^2}{d{\rm ln}\Lambda}=1+\epsilon-c_3L+c_2L+W\frac{c_4}{2}
\ee
\be
\frac{d{\rm ln}S}{d{\rm ln}\Lambda}=\epsilon-2c_3L+2c_2L+Wc_4
\ee
\be
\frac{d{\rm ln}W}{d{\rm ln}\Lambda}=-\epsilon+2c_3L-2c_2L-Wc_4
\ee

The renormalization group flow has a fixed line, as the product
$SW$ is invariant under the RG flow.  It may be verified that the ratio 
$S/W$ has a stable fixed point under RG flow for any $\epsilon$ and
any value of $SW$.  Further, it may be seen that the critical exponent
$\nu$ on the fixed line is given by $\nu d=2$, as if $S$ is constant
under RG flow, then $\sigma_0^2$ has 
$\frac{d{\rm ln}\sigma_0^2}{d{\rm ln}\Lambda}=1+\frac{\epsilon}{2}d/2$.  
Later, we will consider the effect of $1/N$ corrections.

First, note that the line is peculiar to having one time dimension.  For
fewer than one time dimension, there will be a stable fixed point at $W=0$,
which is attractive in the $W$ direction.  Thus, in the framework of
a double-dimensional expansion, one may not see problems at low orders,
as the fixed point has nice behavior for small numbers of time dimensions.
Compare to results in the double-dimensional expansion\cite{dd}.

Further, the presence of the fixed line only required that the renormalization
of the $w(x)\partial_t$ vertex was equal to the renormalization of the
vertex on the left-hand side of the self-consistency equation defining
$\sigma^2$.  This equality will persist to all orders in a loopwise
expansion via a Ward identity.  Thus, we expect that the fixed line is an
exact property of the large $N$ theory.

Let us consider the effect of $1/N$ corrections on this line.  To lowest
order in $1/N$, for weak disorder, the $1/N$ corrections
only modify the naive scaling dimensions in the RG flow.  The scaling dimension
of $\overline\phi\partial_t\phi$ is not changed under $1/N$ corrections.
However, the scaling dimension of $\overline\phi\phi$ is changed by
an amount $\eta=\frac{32}{3\pi^2}\frac{1}{2N}$.  Thus, $1/N$ corrections
will change the RG equations to
\be
\frac{d{\rm ln}S}{d{\rm ln}\Lambda}=2\eta+\epsilon-2c_3L+2c_2L+Wc_4
\ee
\be
\frac{d{\rm ln}W}{d{\rm ln}\Lambda}=-\epsilon+2c_3L-2c_2L-Wc_4
\ee
Then, we find that $SW$ is growing under the RG flow, and the system
goes off to a different fixed point.  The most reasonable guess then is that
in a system with finite $N$ (including
physical systems with $N=1$), the transition is not near the 
MI-SF transition, but is instead of another type, perhaps the phase-only
transition.  Other authors have shown
that, in some cases, the phase-only transition is stable against
weak commensuration effects\cite{herbut}.

\section{Instanton Calculations}
Unfortunately, the ability to carry out instanton calculations in
this system is rather limited.  It will not be possible to calculate
the action for the instanton with any precision, but we will at least
present some arguments about the behavior of the instanton.  The idea
of the calculation is to look for configuration of $w(x)$ and $\sigma^2(x)$
(these configurations are the ``instantons"), such that the self-consistency
equation cannot be solved without including contributions from zero energy
states, as discussed in section 2.  Let us first consider the case in
$2+1$ dimensions.  Let us assume that we try to produce such
states in a region of linear size $L$.
Looking at the lowest energy state in this region, one would expect that
the contribution of spatial gradient terms in the action would lead to 
an energy scale of order $L^{-1}$.  The linear term $w(x)\omega$ in the action
will become important, and produce such a zero energy state,
when $w(x)\omega$ becomes of order $\omega^2$.
This implies occurs when $w(x)$ is of order $\omega$, which implies
$w(x)\approx L^{-1}$.  For some appropriate
configuration of $w(x)$, assuming quadratic fluctuations in $w(x)$ with
strength of order $W$,
we will have an action $S_{\rm instanton}\propto \int \frac{w^2(x)}{W} d^2x$.
Thus, these configurations will occur with exponentially small probability
$e^{-S_{\rm instanton}}$ for weak disorder in $w(x)$.

Away from $2+1$ dimensions, one will find that the action for the
instanton, ignoring fluctuation corrections, is dependent on scale.  For
$d>2$ it is increasing as the scale increases, indicating that large
instantons are not present.  For $d<2$, it is decreasing as the scale
increases, indicating that large instantons are easy to produce.  This is
simply a way of restating the fact that fluctuations in $w(x)$ are, at the pure
fixed point, irrelevant for $d>2$ and relevant for $d<2$.  It is
to be expected that corrections due to fluctuations as considered in
the renormalization group of the previous section will make the action
for the instanton scale invariant.  However, since we do not fully understand
how to calculate instanton corrections even in the simplest $d=2$ case,
the task of combining instanton and fluctuation corrections is 
presently hopeless.
\section{Higher Moments of the Green's Function}
Having considered the weakly commensurate case, and found no fixed
point in physical systems, we return to the strongly commensurate
case with $W=0$, and consider the behavior of higher moments of the Green's 
function.  A lowest order calculation will show log-normal fluctuations in the
Green's function.

Let us first consider the second moment of the Green's function.  That
is, we would like to compute the disorder average of the square of the
Green's function between two points, which we may
write as $\langle G(0,x)^2 \rangle$.  We may Fourier transform the
square to obtain $\langle G^2(p,\omega) \rangle$.  Now, one
may, when averaging over disorder, include terms in which disorder
averages connect the two separate Green's function.  At lowest order,
there is no low-momentum renormalization of the two Green's function propagator,
beyond that due to the renormalization of each Green's function separately.
See figure 6.
That is, if one imagines the two Green's functions entering some diagram,
with both Green's functions at low momentum, going through a sequence of 
scatterings, and exiting, again with both Green's functions at low momentum,
one does not, to lowest order, find any contribution with lines
connecting the two Green's function.  The reason for this is that
at this order we will only join the Green's functions with a single
line, along which one must have momentum transfer of order $\Lambda$.
This then requires that some of the ingoing or outgoing
momenta must be of order $\Lambda$.

One does, however, find a contribution which we may call a renormalization of
the vertex.  See  figure 7.
In order to find the second moment of the Green's function,
one must start both Green's functions at one point, and end both
Green's functions at another point.  Near the point at which both Green's
functions start, one may connect both lines with a single scattering
off of $\lambda$, at high wavevector of order $\Lambda$.  Then, one
can have a large momentum of order $\Lambda$ circulating around the
loop formed, while the two lines that leave to connect to the rest of
the diagram still have low momentum.  This then replaces the two
Green's function vertex, which we will refer to as $V_2$, by a
renormalized vertex.

The result of the above contribution is that the two Green's function 
vertex $V_2$ is renormalized under RG flow as
\be
\frac{d{\rm ln}V_2}{d{\rm ln}\Lambda}=
c_3L=c_3 c_1^2\Lambda^{8-2D}S
\ee
Then, the second moment of the Green's function, at momentum scale $p$
is given in terms of the first moment by
\be
\langle G^2(p,\omega) \rangle \propto \langle G(p,\omega) \rangle^2 p^{-2c_3L}
\ee
Note the factor of 2 in front of $c_3 L$, as the second moment of the
Green's function gets renormalized at both vertices.  One must insert
the value of $L$ at the fixed point into the above equation to
obtain the behavior of the second moment.

For higher moments, the calculation is similar.  In this case, one must,
for the $n$-th moment, renormalize a vertex $V_n$.  The result is
\be
\frac{d{\rm ln}V_n}{d{\rm ln}\Lambda}=
\frac{n(n-1)}{2}c_3L=\frac{n(n-1)}{2}c_3 c_1^2\Lambda^{8-2D}S
\ee
The factor $\frac{n(n-1)}{2}$ arises as at each stage of the RG one may
connect any one of the lines in the vertex $V_n$ to any other line in the
vertex.  There are $\frac{n(n-1)}{2}$ ways to do this.
Then one finds
\be
\langle G^n(p,\omega) \rangle \propto \langle G(p,\omega) \rangle^n 
p^{-n(n-1)c_3L}
\ee

This result for the behavior of the higher moments of the Green's function
is quite typical for disordered systems.  Compare for example to the results
on 2-dimensional Potts models\cite{ludwig}.  From the results for the
moments of the Green's function one may, under mild assumptions,
determine the distribution function of the Green's function.  This distribution
function is the probability that, for a given realization of disorder, the 
Green's function between two points assumes a specific value.  From the result 
for the moments given above one finds that the distribution function is 
log-normal.  That is, the log of the function has Gaussian fluctuations.  
Physically this should be expected from any lowest order calculation, as 
lowest order calculations generally treat momentum scales hierarchically, and 
one is simply finding that at each scale there are random multiplicative
corrections to the Green's function, causing the log of the Green's function
to obey a random walk as length scale is increased.
\section{Glassy Behavior in the Large $N$ Limit}
First, we would like to demonstrate that, in the large $N$ limit, the
self-consistency equation always has a unique solution.  For
generality, we consider here the case of quartic interactions instead
of $\delta$-function interactions.
In the absence of terms linear in $\omega$, uniqueness is clear on physical
grounds, for the models considered above in which the coupling between
neighboring fields $\phi$ is ferromagnetic and unfrustrated.  However,
we will show this to be true for any coupling between neighboring
fields and in the presence of terms linear in $\omega$.  

Note that, for finite $N$, the terms linear in $\omega$ lead to frustration.
Consider an $N=1$ system with a finite number of sites.
Assume that there is no hopping between 
sites, but there is some repulsion between sites due to a quartic term.
Let there be terms linear in $\omega$ in the action, but no terms 
quadratic in $\omega$.  The 
states of the theory are then determined by how many particles occupy
each site.  The repulsion leads to an effective 
anti-ferromagnetic interaction, in the case in which each site has
zero or one particles and we imagine one particle to represent
spin up and no particles to represent spin down.  
This can then produce frustration.  Compare to the Coulomb gap
problem in localized electron systems\cite{efros}.

However, physically speaking, as
$N \rightarrow \infty$, the discreteness of particle number on each
site disappears, and the system becomes unfrustrated.  We will now
show this precisely.

Consider a problem at non-zero temperature, so that there is a
sum over frequencies $\omega$.  Consider an arbitrary single particle
Hamiltonian $H_0$, defined on a $V$-site lattice, so that the
self-consistency equation involves finding $\lambda_i$, where
$i$ ranges from $1$ to $V$, such that 
\be
\label{arbsc}
\sigma_i^2+\sum\limits_{j} M_{ij}\lambda_j
=\sum\limits_{\omega}\langle i|(H_0+\lambda_i+iA_i\omega
+B_i\omega^2)^{-1}|i\rangle
\ee
Here, $\sigma_i^2$ is a function of site $i$, and $A_i$ and $B_i$
are functions of site $i$ defining the local value of the linear
and quadratic terms in the frequency.  The matrix $M_{ij}$ is included
to represent the effects of a quartic interaction.  For the problem
to be physically well defined, $M_{ij}$ must be positive definite.

The proof that equation (\ref{arbsc}) has only one solution
proceeds in two steps.  First we note that if $H_0$ vanishes, then
the equation obviously only has one solution with
$\lambda\geq 0$.  Next, we will show
that as $H_0$ varies, $\lambda_i$ varies smoothly, and therefore
any arbitrary $H_0$ can be deformed smoothly into a vanishing $H_0$,
leading to a unique solution for $\lambda_i$ for arbitrary $H_0$.

Consider small changes $\delta H_0$ and $\delta\lambda_i$.  In order
for the self-consistency equation to remain true, if we define
\be
v_i= -\sum\limits_{\omega}\langle i|
(H_0+\lambda_i+iA_i\omega+B_i\omega^2)^{-1}
\delta H_0
(H_0+\lambda_i+iA_i\omega+B_i\omega^2)^{-1}|i\rangle
\ee
we must have
\be
\label{linop}
v_i=\sum\limits_{\omega}\langle i|
(H_0+\lambda_i+iA_i\omega+B_i\omega^2)^{-1}
\delta\lambda_i
(H_0+\lambda_i+iA_i\omega+B_i\omega^2)^{-1}|i\rangle
+\sum\limits_{j} M_{ij} \lambda_{j}
\ee
The right hand side of equation (\ref{linop}) defines a linear function
on $\delta\lambda_i$.  If it can be shown that this function is invertible,
then the theorem will follow.  
However, we have that
\be
{\rm Tr}(
\delta\lambda_i
\sum\limits_{\omega}
(H_0+\lambda_i+iA_i\omega+B_i\omega^2)^{-1}
\delta\lambda_i
(H_0+\lambda_i+iA_i\omega+B_i\omega^2)^{-1})>0
\ee
due to the well known fact that second order perturbation theory always
reduces the free energy of a quantum mechanical system with a Hermitian
Hamiltonian at finite temperature.
We also have, as discussed above, that $M_{ij}$ is positive definite.
Thus, the linear function on $\delta\lambda_i$ defined above is a sum of
positive definite functions, and hence positive definite.
Therefore, it is invertible and the desired result follows.

This result is interesting considering the phenomenon of replica
symmetry breaking.  It has been noticed by several authors that large
$N$ infinite-range spin glass models do not exhibit replica symmetry breaking
within a meanfield approximation, both in the classical and 
quantum cases\cite{rsb}.  Although those calculations were based on the
absence of unstable directions, in the large $N$ limit, for fluctuations
about the replica symmetric state, it is possible that the real reason
for the absence of replica symmetry breaking is the uniqueness of the
solution of the self-consistency equation, as shown above.

A second interesting question, having begun to consider possible glassy
behavior in the large $N$ limit, has to do with the nature of the ground
state in the classical limit.  If we drop all terms in $\omega$, to
produce a classical problem, and ask for the classical ground state,
for some arbitrary bare Hamiltonian $H_0$, one may ask how many of
the $N$ available spin components will be used.

In this case, consider Hamiltonian $H_0$, which is a $V$-by-$V$ matrix
in the case where there are $V$ sites.  First consider the case in
which $H_0$ is a real Hermitian matrix.  Since we are considering arbitrary
Hamiltonians $H_0$, we can, without loss of generality, constrain all
spins to be the same length.  We can find the classical ground
state by looking for solutions of the self-consistency equation
\be
\sigma^2=\langle i|(H_0+\lambda_i)^{-1}|i\rangle
\ee
in the limit as $\sigma^2\rightarrow\infty$.

In this limit, the right-hand side will be dominated by zero energy
states (more precisely, states that tend to zero energy as $\sigma^2$ tends to
infinity) of the operator $H_0+\lambda$.  If the system has $k$ of these
states, the ground state of the system will use $k$ of the spin components.
If the system {\it needs} to use all $k$ of these components to form
a ground state, that is, ignoring the case in which a state using $k$
spin components is degenerate with a state using fewer components, then
even under small deformations of $H_0$ the system will use $k$ spin
components in the ground state.  Then, under these small deformations,
$H_0+\lambda$ will still have $k$ zero eigenvalues.  To produce $k$
zero eigenvalues for all real Hermitian matrices in a neighborhood
of a given Hermitian matrix $H_0$ requires 
$k(k+1)/2$ free parameters.  The elements of $\lambda$ provide these
parameters.  Since there $V$ of these elements, we find that
$k(k+1)/2\leq V$, and the number of spin components needed to form
the classical ground state is at most $\sqrt{2V}$.  

If $H_0$ were an arbitrary Hermitian matrix, with complex elements, or
a symplectic matrix, one would find a similar result, with $k$ still
at most order $\sqrt{V}$, although the factor $2$ would change.  This
is analogous to the different universality classes in random matrix
theory\cite{rmt}.  Finally, we make one note on the number of parameters
available to solve the self-consistency equation.  There are $V$ free 
parameters.  However, self-consistency requires solving $V$ independent
equations, so the number of variables matches the number of equations.  By
considering the number of parameters requires to produce zero eigenvalues
of $H_0+\lambda$, we were able to obtain a bound on the number of zero
eigenvalues.  Still, one might wonder if there are enough free parameters
to produce multiple zero eigenvalues and still solve the self-consistency
equations, as it appears that one would then need $k(k+1)/2+V$ free
parameters.  However, if there are $k$ zero eigenvalues, by considering the
different ways of populating the zero energy states (that is, considering
the different ways in which the eigenvalues tend towards zero as
$\sigma^2$ tends toward infinity) one obtains an additional $k(k+1)/2$
parameters, so the number of parameters available always matches the number
of equations.

We can extend this theorem to look at metastable states.  Suppose
a configuration of spins is a local extremum of the energy $H_0$, for
fixed length of spins.  Then, since the derivative of the energy vanishes,
one finds that a matrix $(H_0+\lambda_i)$ must have a number
of zero eigenvalues equal to the number of spin components used.
Suppose that for small deformations of $H_0$ there is still a nearby local 
minimum, as one would like to require for a stable state.
Then, one can argue that the number of spin components $k$
used in the state obeys $k(k+1)/2\leq V$.  

This second theorem may be of interested in considering the onset of
replica symmetry breaking.  If we have a system in a large volume and large
$N$ limit, one must ask in which order the limits are taken.
If the $N\rightarrow\infty$ limit is taken first, there will be no
replica symmetry breaking.  However, if the infinite volume limit is
taken first, there may be replica symmetry breaking.  
If one has $N \geq 2k_{\rm max}$, where $k_{\rm max}$ is the largest $k$ such 
that $k(k+1)/2 \leq V$, then there are no local minima other than
the ground state.  This follows because, as shown above, a local extremum of
the energy, $\phi_i$,
will use at most $k_{\rm max}$ spin components, and the ground state,
$\phi^{\rm gr}_i$, can be constructed using a different set of
$k_{\rm max}$ spin compoents.  Then starting from $\phi_i$,
one finds that deforming the state along the path
$\sqrt{1-\delta^2}\phi_i+\delta\phi^{\rm gr}_i$ as $\delta$ goes
from 0 to 1 provides an unstable direction for fluctuations.
\section{Conclusion}
In conclusion, we have considered the large $N$ dirty boson model,
including the effects of local incommensuration (the terms linear in
$\omega$).  In the large $N$ limit, a fixed line under RG is found,
but is destabilized by including $1/N$ corrections.  This suggests that
the phase transition in experimental ($N=1$) systems is of the phase-only
type, instead of the MI-SF type.

There is a problem with local incommensuration in a perturbative
approach, as discussed in the section on the
Bose glass in the large $N$ limit and the section on instanton calculations.
One would like a quantitative method of assessing the results of the
instantons, although this is largely a technical issue, as it appears
that there are no accessible fixed points in the RG using this approach.

In the strongly commensurate case, it has been shown that one can 
calculate higher moments of the correlation functions.  The result shows 
that the correlation functions have a log-normal distribution.

The two theorems proved in the last section give useful information
on the relevance of the large $N$ expansion in frustrated
problems.  It would be interesting to use these results as a starting
point for a better understanding of replica symmetry breaking.

The large $N$ approximation has been a useful approximation for pure,
unfrustrated systems.  It is hoped that it may become as useful for
disordered interacting systems.
\section{Acknowledgements}
The result for the number of spin components needed
to form a ground state was obtained in collaboration with David Huse,
who I would also like to thank for many useful discussions on other
results in this work.  I would like to thank the ICTP, in Trieste, Italy,
for their hospitality while some of this work was in progress.

\newpage
\begin{figure}[!t]
\begin{center}
\leavevmode
\epsfig{figure=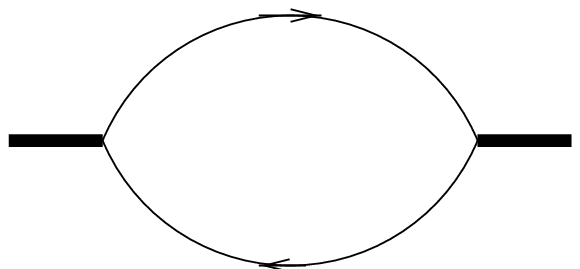,height=3cm,angle=0}
\end{center}
\caption{Polarization bubble.  Thick lines represent either scattering vertex
off $\lambda$ or scattering vertex used to define $\sigma^2$ in
self-consistency equation.}
\label{fig1}
\end{figure}
\begin{figure}[!t]
\begin{center}
\leavevmode
\epsfig{figure=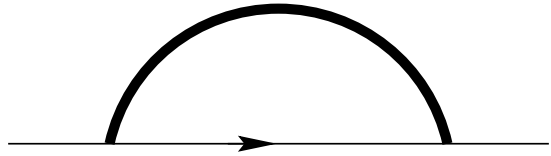,height=2cm,angle=0}
\end{center}
\caption{Self-energy correction due to fluctuations in $\lambda$.  Joining the
thick lines in a loop denotes averaging $\lambda$ over disorder in 
$\sigma^2$. Momentum of order $\Lambda$ flows around loop.}
\label{fig2}
\end{figure}
\begin{figure}[!t]
\begin{center}
\leavevmode
\epsfig{figure=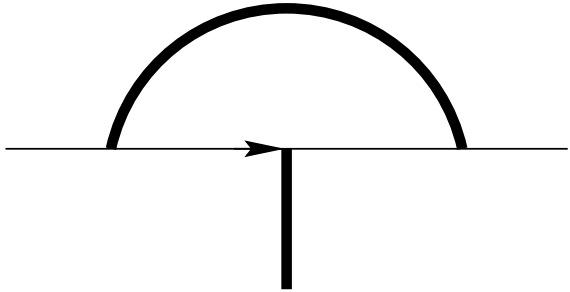,height=4cm,angle=0}
\end{center}
\caption{Vertex correction due to fluctuations in $\lambda$.  This represents
both renormalization of vertex defining scattering off of $\lambda$ and
renormalization of vertex defining $\sigma^2$ in self-consistency equation.}
\label{fig3}
\end{figure}
\newpage
\begin{figure}[!t]
\begin{center}
\leavevmode
\epsfig{figure=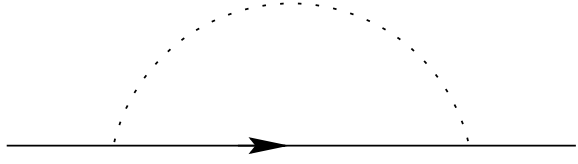,height=2cm,angle=0}
\end{center}
\caption{Renormalization of $\omega^2$ term due to fluctuations
in $w(x)$.  The vertices  with thin lines represent scattering off of 
$w(x)\partial_t$, while the joining of the vertices represents averaging
$w(x)$ over disorder.}
\label{fig4}
\end{figure}
\begin{figure}[!t]
\begin{center}
\leavevmode
\epsfig{figure=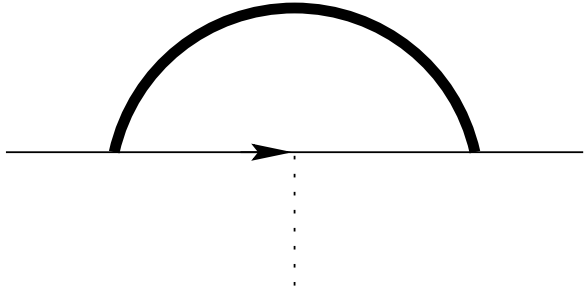,height=4cm,angle=0}
\end{center}
\caption{Renormalization of vertex for $w(x)$ due to fluctuations in
$\lambda$.}
\label{fig5}
\end{figure}
\begin{figure}[!t]
\begin{center}
\leavevmode
\epsfig{figure=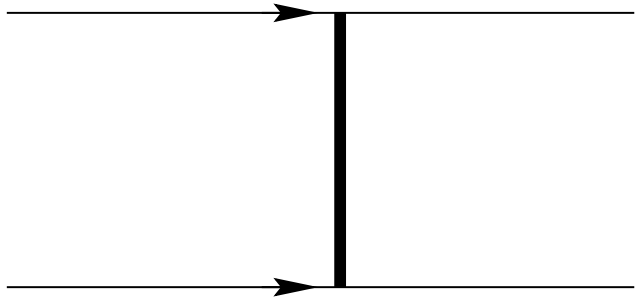,height=4cm,angle=0}
\end{center}
\caption{Possible contribution to propagation of two Green's functions,
both in same realization of disorder.  This is used to compute second moment 
of the Green's function.  This diagram does {\it not} lead to any 
renormalization of low momentum behavior.}
\label{fig6}
\end{figure}
\newpage
\begin{figure}[!t]
\begin{center}
\leavevmode
\epsfig{figure=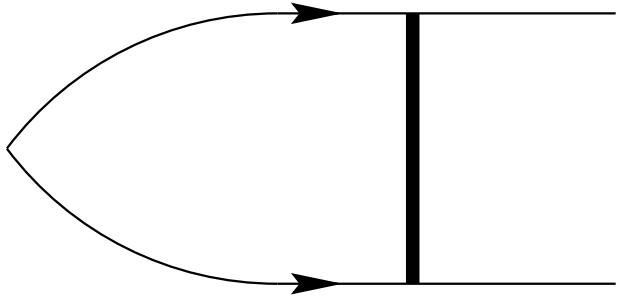,height=4cm,angle=0}
\end{center}
\caption{Renormalization of vertex in computing higher moments of
Green's function.  Two Green's functions start at the same point.  After
Fourier transforming, this implies that they start with given
total momentum.  By including fluctuations in $\lambda$, with momentum 
of order $\Lambda$ running around the loop, one can define a renormalized 
vertex.}
\label{fig7}
\end{figure}
\end{document}